# Transformation of equations in analysis of proportionality through referent models.

Enrique Ordaz Romay[1]

*Facultad de Ciencias Físicas, Universidad Complutense de Madrid*

## Abstract

In proportionality of objects, samples or populations, usually we work with *Z* score of proportionality calculated through referent models, instead directly with the variables of the objects in itself. In these studies we have the necessity to transform, the equations that use the variables of the object, in equations that directly use like variables *Z* score.

In the present work a method is developed to transform the parametric equations, in equations in variables *Z* using like example the studies of human proportionality from the Phantom stratagem of Ross and Wilson.

## Introduction

The study of the proportionality of objects, samples or populations usually is approached from the analysis of *Z* score calculated from a model referent, from the expression:

$$Z_{variable_1, variable_2; P}(i) = \frac{variable_1(i)\left(\dfrac{variable_2(P)}{variable_2(i)}\right)^n - variable_1(P)}{s_{variable_1}(P)} \tag{1}$$

being *i* the subject of study, *P* the referent model, $variable_1$ the study variable, $variable_2$ the proportionality base, $n = n_1/n_2$ being $n_1$ and $n_2$ the dimensions of $variable_1$ and $variable_2$ respectively and $s_{variable_1}$ the standard deviation of $variable_1$.

---

[1] eorgazro@cofis.es

The referent model will be an ideal model of object that is used like standard of proportionality measurement. Times it will be a model formed by statistical averages, other times a comfortable mathematical model or useful to even facilitate the calculations and in some cases, the models will be established like canons of some specific quality.

Let us suppose that we used, as standard, the parameters of a sample. In these conditions, if the distribution of the variables of the sample is a normal distribution, referent will be defined by a table that contains the values of the averages and standard deviations of the different variables from the sample. That is to say, referent model $P$ will be defined by table 1.

| Variable | Average | Deviation Standard |
|---|---|---|
| $Variable_1$ | $Average_1$ | $Devstd_1$ |
| $Variable_2$ | $Average_2$ | $Devstd_2$ |
| $Variable_3$ | $Average_3$ | $Devstd_3$ |
| … | … | … |
| $Variable_m$ | $Average_m$ | $Devstd_m$ |

Table 1. Prototype of table of $P$ referent

For example: In the studies of human proportionality, Ross and Wilson in 1974 defined a proportionality standard to which they called "Phantom" model so that it served like reference. The calculation of the values corresponding to the kinanthropometric variables of the Phantom model was based on extensive data bases of general population. The perimeters were obtained from the data base of Wilmore and Behnke in 1969 and 1970, the skinfolds from a data base not published of Yuhasz (Carter, 1996) and the rest of measures was obtained by Garret and the Kennedys in 1971. The result of these works lead until the table of the Phantom reference model.

| Variable | Average | Deviation Standard |
|---|---|---|
| Height | 170.18 cm | 6.29 cm |
| Acromial height | 139.37 cm | 5.45 cm |
| Radial height | 107.25 cm | 5.37 cm |
| … | … | … |
| Weight | 64.58 kg | 8.60 kg |

Table 2. Fragment of the table of the human model "Phantom".

In order to calculate *Z* score, Ross and Wilson Phantom method uses the equation:

$$Z_{variable;Ph}(i) = \frac{variable(i)\left(\frac{height(Ph)}{height(i)}\right)^n - variable(Ph)}{s_{variable}(Ph)} \quad (2)$$

That it is deduced from equation (1) replacing variable$_2$ by stature and the referent model *P* by the Phantom symbolized by *Ph*. Like, to use the stature as it bases of the proportionality, is inherent part of the Phantom method gets used to omitting this subscript.

## Parametric equations and *Z* score equations.

We will call parametric equation to that equation whose result is obtained by direct substitution of the study object values in the variables of the equation. Its traditional form, being $x_1(i), x_2(i), …, x_m(i)$ the values of our study object *i* (object o sample), will be:

$$f(x_1(i), x_2(i), …, x_m(i))$$

For example, in kinanthropometry, for the calculation of the fat mass percentage, one of the equations that are used is:

- Equation of Faulkner to calculate the fat mass, whose expression has the form (Sk. = Skin-fold and the skin-folds are measured in *mm*):

$$\%Fat\ mass = 0.153 \cdot (Sk.triceps + Sk.subscapula + Sk.suprailiac + Sk.abdominal) + 5.783$$

In this equation, they are used, like variables for the calculation of the percentage of fat mass, four values directly measured an individual *i* (or, in a sample, the average of these variables). Other equations of the same type that we will use like examples, each one because it allows to study a quality different from the method, will be the following ones:

- Percentage of the fat mass of Yuhasz (just as before, the sum is of skin-folds and the they are measured in *mm*):

$$\%Fat\ mass(i)_{female} = 4.56 + \sum_{v=triceps,subesc.,suprail.,abdom.,thigh,leg} 0.143 \cdot v(i)$$

$$\%Fat\ mass(i)_{male} = 3.64 + \sum_{v=triceps,subesc.,suprail.,abdom.,thigh,leg} 0.097 \cdot v(i)$$

- Bony mass of Von Döbeln modified by Rocha (*d.* = diameter, the units is *kg* and *m*):

$$Bony\ mass = 3.023 \cdot (400 \cdot heigh^2 \cdot d.styloid \cdot d.bicondileo\ femoral)^{0.712}$$

On the other hand, we will call *Z* score equation to any equation whose result is obtained by substitution (not of the direct values) of *Z* score of the study object proportionality in the variables of the equation. To this *Z* score function will have the form:

$$f\left(Z_{x_1}(i), Z_{x_2}(i), ..., Z_{x_m}(i)\right)$$

The equations of the last type are not easily in the bibliography, but they are not difficult to deduce. Let us see the process exemplifying it with the calculation of the ponderal index (*PI*) based on its *Z* scores. The equation of the *PI* is:

$$PI(i) = \frac{height(i)}{\sqrt[3]{weight(i)}} \tag{3}$$

That is to say, *PI* is, traditionally and by definition, a parametric function of two direct variables: *height* and *weight*.

The form to obtain the equation in *Z* score from the parametric equation (3) is, outline, simple. In the first place we cleared *variable*(*i*) of the equation (2) obtaining:

$$variable(i) = \left(variable(Ph) + Z_{variable;Ph}(i) \cdot s_{variable}(Ph)\right)\left(\frac{height(i)}{height(Ph)}\right)^n \tag{4}$$

Replacing this expression in the parametric equation we obtain as result the equation in *Z* score. In the case of *PI* the equation (4), for *variable* = *weight* and knowing that, in the Phantom model of human proportionality *weight*(*Ph*) = 64,58 *kg*, $s_{weight}(Ph)$ = 8,60 *kg* and that the *weight* has dimension *n* = 3 is:

$$weight(i) = \left(64.58\ kg + 8.60\ kg \cdot Z_{weight;Ph}(i)\right)\left(\frac{height(i)}{170.18\ m}\right)^3$$

We only must replace this expression in the parametric equation of IP (3) for obtain the *PI* equation in *Z* scores:

$$PI(i) = \frac{170.18}{\sqrt[3]{8.60 \cdot Z_{weight,Ph}(i) + 64.58}}$$

This is the looked equation, that is to say, the formula that calculates the ponderal index based on *Z* score of the weight.

## Simplification of the *Z* scores equation.

To replace the equation (4) in the expression of the parametric equation for obtain the equation in *Z* scores, normally leads to complex expressions that do not respond exactly with the necessities which arise when we considered the transformation of equations in the analysis of the proportionality.

The circumstances in which the search of the equation in *Z* scores takes place are the following ones:

- A form simple is needed (preferably linear) to evaluate, by means of *Z* scores, a parameter that traditionally calculates through a parametric expression.

- We know that, when *Z* scores of the variables involved in the parametric equation are zero, the parameter calculated with this equation agrees proportionally with the value of those same parameters in the referent model.

- We also know that, in general, the referent models (universal or particular) are used in proportionality so that *Z* score has small values near zero. In fact, great values for *Z* score, more than disproportions in the study objects, indicate in many occasions, the inadequate of the referent model to this concrete study.

Evidently, the terms "small" and "great" are very subjective and will depend on the objectives of the project or study. Nevertheless, generally, these terms can be done objectives from the set of the work of investigation, using the significance level (*p*-value) of the study, that is to say, the precision of the study. The form consists of defining "small" *Z* score when.[2]

$$Z_{variable} < \frac{variable(P)}{s_{variable}(P)} + ln(p) \qquad (5)$$

For example, using like referent the model of human proportionality Phantom, for the variable *weight* ($v_{weight}(Ph) = 64.58$ *kg* and $s_{weight}(Ph) = 8.60$ *kg*) with $p < 0.01$, the acceptable value for $Z_{peso}$ must be smaller than $^{64.59}/_{8.60} + ln(0,01) \approx 2.90$. That is to say, if $Z_{weight}(i)$ score, in one individual *i*, is smaller than 2.90 the Phantom model of human proportionality can be applied to this individual *i* with a significance level $p < 0.01$.

---

[2] An estimation of *Z* takes form $Z = Z_{estimate} - \varepsilon$, being ε the relative error that is committed with the estimation. In first approach, the precision *p* is proportional to the exponent of the minus relative error ($p \sim e^{-\varepsilon}$) and *Z* must be of the order or smaller than *variable*(*P*)/*s*(*P*), replacing and clearing $Z_{estimate}$ the expression (5) is obtained.

Thanks to this two circumstances that concur in the calculation of Z score we can make a linear approach to a function $f(Z_{x_1}(i), Z_{x_2}(i), ..., Z_{x_m}(i))$

Replacing (4) in $f(x_1(i), x_2(i), ..., x_m(i))$ we found:

$$f(..., Z_{x_j}(i),...) = f\left(..., \left(x_j(Ph) + Z_{x_j;Ph}(i) \cdot s_{x_j}(Ph)\right)\left(\frac{height(i)}{height(Ph)}\right)^{n_j}, ...\right)$$

Taking as variable Z score values and developing this equation by McLaurin series, we will have:

$$f(..., Z_{x_j}(i),...) = f(..., Z_{x_j}(i),...)\bigg|_{Z=0} + \sum_j \frac{\partial f}{\partial Z_{x_j}}\bigg|_{Z=0} \cdot Z_{x_j}(i) + o(Z_{x_j}(i)) \qquad (6)$$

Let us calculate each term of this expression: In the first place, evaluate the function in Z scores in the value $Z = 0$ is just like to evaluate the parametric equation in the point of the referent one. By virtue of (4) we can write: $f(..., Z_{x_j}(i),...)\bigg|_{Z=0} = f(Ph) \cdot \left(\frac{height(i)}{height(Ph)}\right)^n$ being $n$ the dimension of the $f$ parameter.

Since, according to the laws of the implicit functions: $\frac{\partial f}{\partial Z_x} = \frac{\partial f}{\partial x} \cdot \frac{\partial x}{\partial Z_x}$ and being: $\frac{\partial x}{\partial Z_x} = s_x(Ph) \cdot \left(\frac{height(i)}{height(Ph)}\right)^{n_x}$ that evaluated in $Z = 0$ (remember that $Z = 0$ does not force that $height(i) = height(Phantom)$ and therefore the base of the exponent does not become one), and $\frac{\partial f}{\partial x}\bigg|_{Z=0} = \frac{\partial f}{\partial x}(Ph) \cdot \left(\frac{height(i)}{height(Ph)}\right)^{n-n_x}$, consequently the derivative is:

$$\frac{\partial f}{\partial Z_{x_j}}\bigg|_{Z=0} = s_x(Ph) \cdot \frac{\partial f}{\partial x}(Ph) \cdot \left(\frac{height(i)}{height(Ph)}\right)^n.$$

Replacing in (6) we have left:

$$f(...,Z_{x_j}(i),...) \approx f(Ph) \cdot \left(\frac{height(i)}{height(Ph)}\right)^n + \sum_j \left(\frac{\partial f}{\partial x_j}(Ph) \cdot s_{x_j}(Ph) \cdot Z_{x_j}(i) \cdot \left(\frac{height(i)}{height(Ph)}\right)^n\right)$$

Making $C_{f,j} = \frac{\partial f}{\partial x_j}(Ph)$ and comparing with (4) we can say that exist a value *m* for which this last expression can be rewritten in the form:

$$f(...,Z_{x_j}(i),...) \approx \left(f(Ph) + \sum_j \left(C_{f,j} \cdot s_{x_j}(Ph) \cdot Z_{x_j}(i)\right)\right) \cdot \left(\frac{height(i)}{height(Ph)}\right)^n \quad (7)$$

Thus, (7) will be the equation looked for the calculation of parameter f from Z score. We will call to this equation SZSE (simplified in Z score equation). The values of the $C_{f,x}$ coefficients can easily be found by differential calculus for each equation or be deduced by numerical analysis. For *f* equations that are linear in their variables (as those of Faulkner or Yuhasz), the coefficients that multiplies to the variables in the parametric equation agree with these coefficients.

The physical meaning of the equation (7) is immediate:

- If all Z scores of the variables that take part in the calculation of a parameter are equal to zero, such parameter adopts the proportional value of the referent model.

- A value of an Z score different from zero increases or diminishes the parameter respect the referent value, with a "weight" of $C_{f,x} \cdot s_x(Ph)$.

- In general, in the Z score equation, the base of the proportionality (the stature in the Phantom model), takes part in the calculation of the parameter although it does not appear in the parametric equation.

# Examples of the simplified *Z* score equations.

## 1. Simplified *Z* score equations for ponderal index *PI*.

The equation of the ponderal index is: $PI(i) = \dfrac{height(i)}{\sqrt[3]{weight(i)}}$.

The variables that take part in the calculation of *PI* are *height* and *weight*, therefore, applying the equation (7) we obtain:

$$PI(i) = \left(PI(Ph) + C_{PI,weight} \cdot s_{weight}(Ph) \cdot Z_{weight}(i) + C_{PI,height} \cdot s_{height}(Ph) \cdot Z_{height}(i)\right) \left(\dfrac{height(i)}{height(Ph)}\right)^n$$

It is simple to calculate these values: the parameter *PI* and their equation are dimensionless, therefore *n* = 0. Also a simple substitution makes us see that in the Phantom method of human proportionality $Z_{height}$ score is always equal to zero.

By substitution of values we calculated $PI(Ph) = 170.18/(64.58)^{1/3} \approx 42.41$ and with a consultation to the table of the Phantom model is obtained $s_{weight}(Ph) = 8.60$ *cm*.

Finally, $C_{weight}$ is calculates, by differential calculus, knowing that $C_x = \left.\dfrac{\partial f}{\partial x}\right|_{Ph}$. In this case $C_{weight} = \left.\dfrac{\partial IP}{\partial weight}\right|_{Ph} = \left.\dfrac{\partial}{\partial weight}\left(\dfrac{height}{\sqrt[3]{weight}}\right)\right|_{Ph} = \left.-\dfrac{height}{3 \cdot weight \cdot \sqrt[3]{weight}}\right|_{Ph} \approx -0.2189$

Replacing all the obtained values we found:

$PI(i) \approx 42.41 + (-0.2189) \cdot 8{,}60 \cdot Z_{weight}(i) \rightarrow \boxed{PI(i) \approx 42.41 - 1.88 \cdot Z_{weight}(i)}$

In order to see the differences between these two equations we calculated the minimum *p*-value of a study for different values from stature obtaining table 3.

| *p*-value | Height (*cm*) | | | | | | |
|---|---|---|---|---|---|---|---|
| Weight (*kg*) | 140 | 150 | 160 | 170 | 180 | 190 | 200 |
| 35 | 0.0005 | 0.0110 | 0.0321 | 0.0584 | 0.0864 | 0.1145 | 0.1416 |
| 45 | 0.0128 | 0.0004 | 0.0066 | 0.0236 | 0.0463 | 0.0715 | 0.0975 |
| 55 | 0.0499 | 0.0121 | 0.0004 | 0.0054 | 0.0203 | 0.0408 | 0.0640 |
| 65 | 0.1039 | 0.0403 | 0.0092 | 0.0003 | 0.0057 | 0.0198 | 0.0389 |
| 75 | 0.1684 | 0.0806 | 0.0298 | 0.0059 | 0.0004 | 0.0069 | 0.0210 |
| 85 | 0.2389 | 0.1294 | 0.0593 | 0.0200 | 0.0029 | 0.0010 | 0.0091 |
| 95 | 0.3117 | 0.1838 | 0.0958 | 0.0412 | 0.0120 | 0.0010 | 0.0024 |
| 105 | 0.3840 | 0.2416 | 0.1374 | 0.0680 | 0.0266 | 0.0061 | 0.0003 |

Table 3. Values of minimum *p*-value of the study for the simplified Z score equation of the ponderal index *PI*.

The graphical representation of this table is the graphic 1:

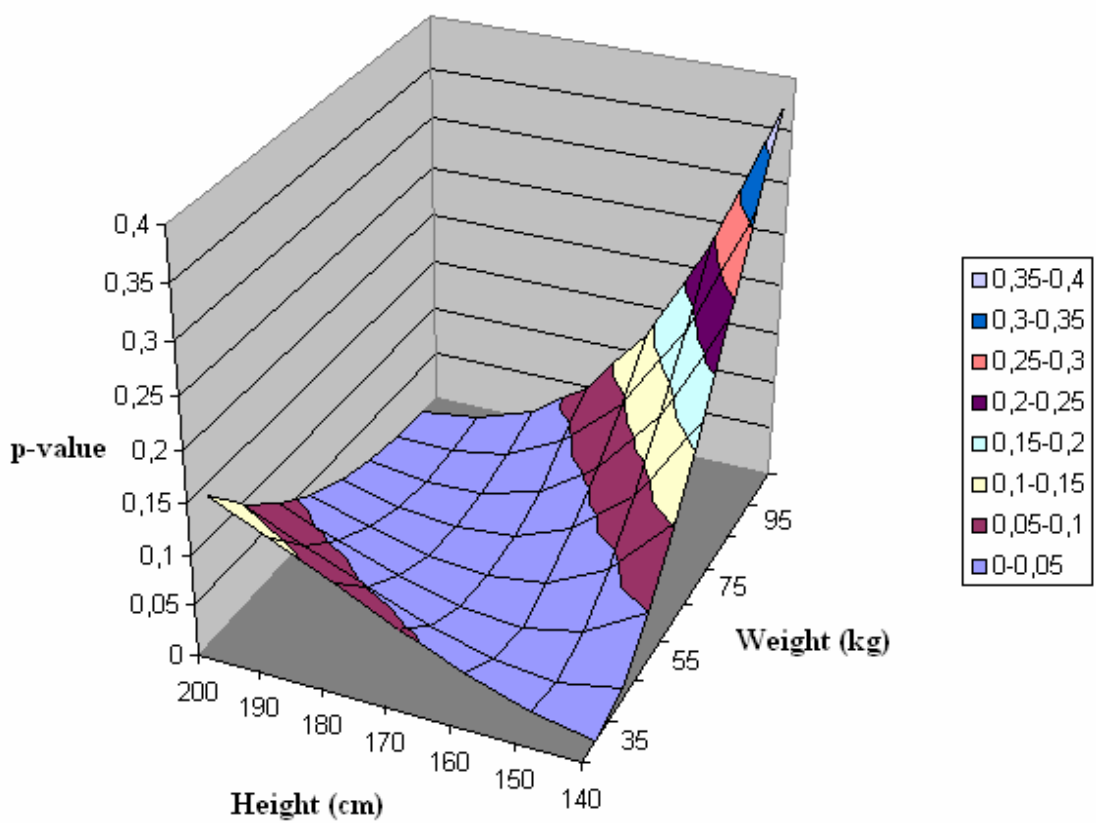

Graphic 1. Values of minimum *p*-value of the study for the simplified Z score equation of the ponderal index *PI*.

From table 3 and graphic 1 we deduced that, the simplification for the equation in *Z* score is fit for values than they stay in concrete surroundings. The study of $Z_{weight}$ score for these same values gives like result table 4 and graphic 2.

| $Z_{weight}$ | Height (*cm*) | | | | | | |
|---|---|---|---|---|---|---|---|
| Weight (*kg*) | 140 | 150 | 160 | 170 | 180 | 190 | 200 |
| 35 | -2.6 | -2.9 | -3.2 | -3.4 | -3.7 | -3.9 | -4.0 |
| 45 | -1.1 | -1.6 | -1.9 | -2.3 | -2.6 | -2.8 | -3.1 |
| 55 | 0.3 | -0.3 | -0.7 | -1.1 | -1.5 | -1.8 | -2.1 |
| 65 | 1.7 | 1.1 | 0.5 | 0.1 | -0.4 | -0.7 | -1.1 |
| 75 | 3.1 | 2.4 | 1.8 | 1.2 | 0.7 | 0.3 | -0.1 |
| 85 | 4.5 | 3.7 | 3.0 | 2.4 | 1.8 | 1.3 | 0.9 |
| 95 | 5.9 | 5.0 | 4.2 | 3.5 | 2.9 | 2.4 | 1.9 |
| 105 | 7.3 | 6.3 | 5.5 | 4.7 | 4.0 | 3.4 | 2.9 |

Tabla 4. Values of $Z_{peso}$.

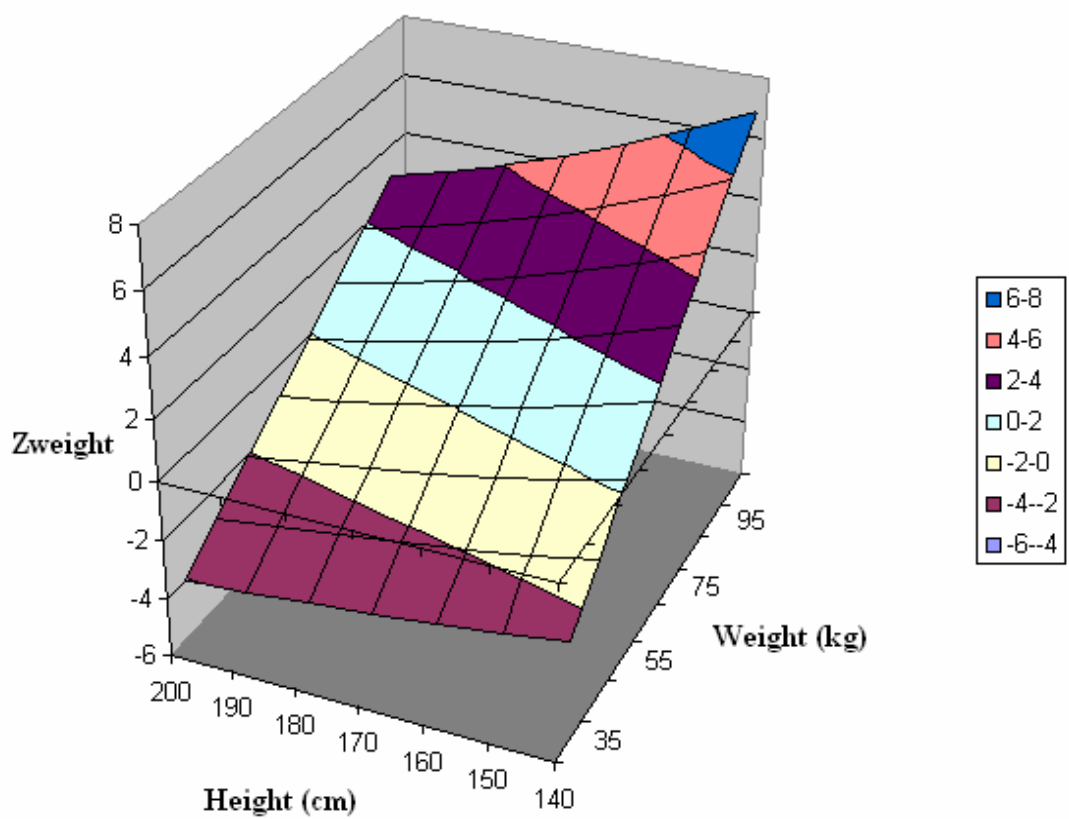

Gráfica 2. Valores de $Z_{peso}$.

It is easy to verify that, indeed, for values of height small and high weight the $Z_{weight}$ indices are great and let fulfill the condition (5). For this reason, the values that better fit both equations (parametric and SZSE) are, for this particular case those that they fulfill $|Z_{weight}| < 4$.

In summary, we have two equations for the ponderal index:

- Parametric equation: $PI(i) = \dfrac{height(i)}{\sqrt[3]{weight(i)}}$

- Z score equation: $PI(Z(i)) \approx 42.41 - 1.88 \cdot Z_{weight}(i)$

The first equation is the parametric traditional one, the second one is used in proportionality when $|Z_{weight}| < 4$.

## 2. SZSE for Faulkner fat mass percentage.

The equation of Faulkner fat mass percentage is a lineal equation in all of its variables and the coefficient that multiplies them is the same one for all of them: 0.153.

Let us calculate the values to replace in the equation (7): the percentage of fat mass is a dimensionless parameter, that is to say, $n = 0$. Also, the percentage of fat mass of the Phantom model (calculated using the formula of Faulkner) is 18.79%. Finally, all the $C_{Faulkner,x}$ take the value from 0.153 in all the variable. Consequently, the equation (7) we have left:

$$\%Fat\ mass(i) = 18.79 + \sum_{\substack{x=triceps,subesc.\\suprail.,abdom.}} 0.153 \cdot s_x(Phantom) \cdot Z_x(i)$$

Proposing an example: If an individual $i$ has a *height* of 168 *cm* and values of $Z_{Triceps}(i) = -1.2$ ; $Z_{Subscapula}(i) = 0.1$ ; $Z_{Suprailiac}(i) = -0.4$ and $Z_{Abdominal}(i) = 0.2$, the calculation of its percentage of fat mass will be the corresponding one to table 5 for (4).

| Variable (skin-fold) | $C_{Faulkner,x}$ | $s_x(Ph)$ | $Z_x(i)$ | Product |
|---|---|---|---|---|
| Triceps | 0,153 | 4,47 | -1,2 | -0,820692 |
| Subscapula | 0,153 | 5,07 | 0,1 | 0,077571 |
| Suprailiac | 0,153 | 4,47 | -0,4 | -0,273564 |
| Abdominal | 0,153 | 7,78 | 0,2 | 0,238068 |
| $Z_{\%comp.} \cdot s_{\%comp.}$ → | | | **Sum** | -0,778617 |

Table 5 Calculation of sum $C \cdot s \cdot Z$.

Applying the equation (3) with $n = -1$ the result is:

$$\%Masa\ grasa(i) \approx 18{,}79 - 0{,}78 = 18{,}01\ \%$$

It agrees to observe that the difference, between the results of the fat mass of the Faulkner parametric equation and the *Z* score equation is zero when the height of *i* is the one of *Phantom*. Nevertheless, although in the Faulkner parametric equation of the height does not take part, it takes part in the SZSE from Faulkner one (though *Z* scores). This fact is excellent because it corrects one of the defects of the Faulkner equation: It is anatomically evident that if two individuals have equal skin-fold, the taller will have a smaller percentage of the fat than the lowest. This data is not had in account in the Faulkner equation, but it is contemplated in the equation derived from the proportionality because in the taller the *Z* sores are smaller than the lowers then the skin-folds are equal in both.

## 3. SZSE for Yuhasz fat mass percentage

Very similar to the Faulkner equations are the equations of the percentage of Yuhasz fat mass. These differentiate between men and women, include two skin-fold more (thigh and leg) and changes the coefficients of the variables. Nevertheless, the equations are linear, too. In this form, the two SZSE of the percentage of Yuhasz fat mass, are:

$$\%M.grasa(i)_{femenino} = 18.79 + \sum_{\substack{V=triceps, subscap.\\ suprail., abdom.\\ thigh, leg}} 0.143 \cdot s_V(Phantom) \cdot Z_V(i)$$

$$\%M.grasa(i)_{masculino} = 18.79 + \sum_{\substack{V=triceps, subscap.\\ suprail., abdom.\\ thigh, leg}} 0.097 \cdot s_V(Phantom) \cdot Z_V(i)$$

## 4. SZSE for Von Döbeln bony mass modified by Rocha.

Another example would be the calculation of the Von Döbeln bony mass equation of modified by Rocha. In the first place we will calculate the value of the explaining n for the equation (9): The bony mass, like all the masses, has *dimension* = 3. Nevertheless, the equation of Von Döbeln is not exactly 3, but it has the product of three variables with *dimension* = 1, one of them elevated to the square, and all with exponent 0.712. This corresponds to a dimension $(2+1+1) \cdot 0.712 = 2.848$, value very near, although different from 3. That is to say, the dimensions of the bony mass are different by parameter and by equation[3]. Therefore, $n$ will be an intermediate value between both. We will have then: $2.848 < n < 3$. As election we take the average from these two values: $(3+2.848)/2 = 2.924$.

On the other hand, *Coefficient*$_{height}$ is not necessary to calculate it, because we know that $Z_{height}$ in the Phantom method is always equal to zero. This does not mean that the *height* does not contribute to the bony mass in the SZSE derived from the one from Von Döbeln, but that its contribution is including in being of the explaining *n*.

The rest of coefficients are calculated by analysis differential and their results are *Coefficient*$_{d.styloid}$ = 1.4341 and *Coefficient*$_{d.bicondileo}$ = 0.7848. Multiplying these values by the respective standard deviations we calculated the numbers that multiply to Z scores. Like $s_{d.styloid} = 0.28$ *cm* and $s_{d.bicondileo} = 0.48$ *cm*, the equation we have left:

$$Bony\ mass(i) = \left(10.49 + 0.4015 \cdot Z_{d.styloid} + 0.3767 \cdot Z_{d.bicondileo}\right)\left(\frac{height(i)}{170.18}\right)^{2.924}$$

Applying it to a example: Let an individual $i$ with *height* = 1.88 m, who has the values: $Z_{d.styloid}(i) = 1.3$ ; $Z_{d.bicondileo\ femur}(i) = 1.7$. Its Von Döbeln bony mass will be (table 6):

| Variable (diameter) | Parameter (Döbeln) | $Z_{Var}(i)$ | Product |
|---|---|---|---|
| Styloid | 0.4015 | 1.3 | 0.52195 |
| Bicondileo femur | 0.3767 | 1.7 | 0.64039 |
| $Z_{bony\ mass}(i) \rightarrow$ | | **Sum** | 1.16234 |

Table 6. Calculation of sum $C \cdot s \cdot Z$.

Using the equation (7) the result is:

---

[3] It could seem that in the case of Faulkner or Yuhasz formulas it happen the same, nevertheless, in both cases the equations have an independent term that "assumes" the dimensional difference and for that reason it facilitates that such equations depend on the height through Z scores.

$$\textit{Bony mass}(i) \approx (10.49 + 1.16)\cdot(^{188}/_{170.18})^{2.924} = 15.59 \text{ Kg}$$

It is possible to emphasize that the difference between using the equation of Von Döbeln and the derived one from indices Z is inferior to 1% when Z scores are included in the interval (-4, 4).

# Bibliography.